\documentclass[pre,aps,preprint]{revtex4}
\usepackage{natbib}
\usepackage[cp850]{inputenc}
\usepackage{amsmath,amssymb,color}
\usepackage{graphicx}
\usepackage{latexsym}

\newcommand{\be}{\begin{equation}}
\newcommand{\ee}{\end{equation}}

\newcommand{\fr}{\displaystyle\frac}

\begin{document}

\title{Unconditional jetting}
\author{Alfonso M. Ga\~{n}\'an-Calvo}
\affiliation{E.S.I,
Universidad de Sevilla.\\
Camino de los Descubrimientos s/n 41092 Spain.}
\date{\today}
\begin{abstract}
Capillary jetting of a fluid dispersed into another immiscible phase is usually limited
by a critical Capillary number, a function of the Reynolds number and the fluid
properties ratios. Critical conditions are set when the minimum spreading velocity of
small perturbations $v^*_-$ along the jet (marginal stability velocity) is zero. Here we
identify and describe parametrical regions of high technological relevance, where $v^*_-
> 0$ and the jet flow is always supercritical independently of the dispersed liquid flow rate:
within these relatively broad regions, the jet does not undergo the usual dripping-jetting
transition, so that either the jet can be made arbitrarily thin (yielding droplets of any
imaginably small size), or the issued flow rate can be made arbitrarily small. In this work, we
provide illustrative analytical studies of asymptotic cases for both negligible and dominant
inertia forces. In this latter case, requiring a non-zero jet surface velocity, axisymmetric
perturbation waves ``surf'' downstream for all given wave numbers while the liquid bulk can remain
static. In the former case (implying small Reynolds flow) we found that the jet profile small slope
is limited by a critical value; different published experiments support our predictions.
\end{abstract}


\maketitle

\section{Introductory remarks}

The quest for the conditions under which a given stream of fluid 1 can be dispersed as very small,
homogeneously sized droplets into another immiscible fluid 2 is an old endeavor. Steady capillary
jetting produces droplets of any desired diameter at a controllable rate through Rayleigh-Plateau
instability, and thus is the preferred choice in many applications. Jetting can be supported by a
diversity of energy sources, from plain pressure \cite{Basaran02} or electrostatic suction
\cite{Zel17} (or their combination \cite{eFF06,Gan07}), to chemical potential \cite{JFHomsy04} or
even thermal gradients. Very recently, jetting has also been shown to take place under concentrated
photon irradiation (laser) when surface tension is extremely low \cite{Casner2003} or even under
ultrasound irradiation by the same group. Capillary jetting from a fluid source gives rise to
droplets smaller than dripping (a phenomenon where drops are individually issued from the source at
a certain frequency) under the same operating conditions. Consequently, a significant effort has
been lately devoted to map the transition from jetting to dripping from a fluid source
\citep{Lin2003,Sev05,AP06,Guillot2007}, in the search for extended jetting conditions down to the
smallest possible jet diameter. Two key dimensionless numbers gauge the role of inertia and viscous
forces relative to surface tension, namely Weber and Capillary numbers $W\!e=\rho_1 U_s^2 R/\sigma$
and $Ca=\mu_1 U_s/\sigma$, $\rho_1$, $\mu_1$, $\sigma$ being the density, viscosity and surface
tension of the dispersed fluid. $U_s$ and $R$ are the jet surface velocity and radius,
respectively. Alternatively, the Reynolds number ${\text Re}=W\!e/Ca=\rho_1 U_s d/\mu_1$ and $Ca$
can be used to characterize the jet dynamics; this choice is particularly useful in microfluidics,
where ${\text Re}$ is usually moderate or small (laminar flows). Surface tension is the main agent
sustaining wave propagation of disturbances along the jet (downstream coordinate $z$), and thus
$Ca$ is the key parameter controlling the jet dynamics in microfluidics when ${\text Re}$ is small.
\begin{figure} \centerline{\includegraphics[width=0.6\textwidth]{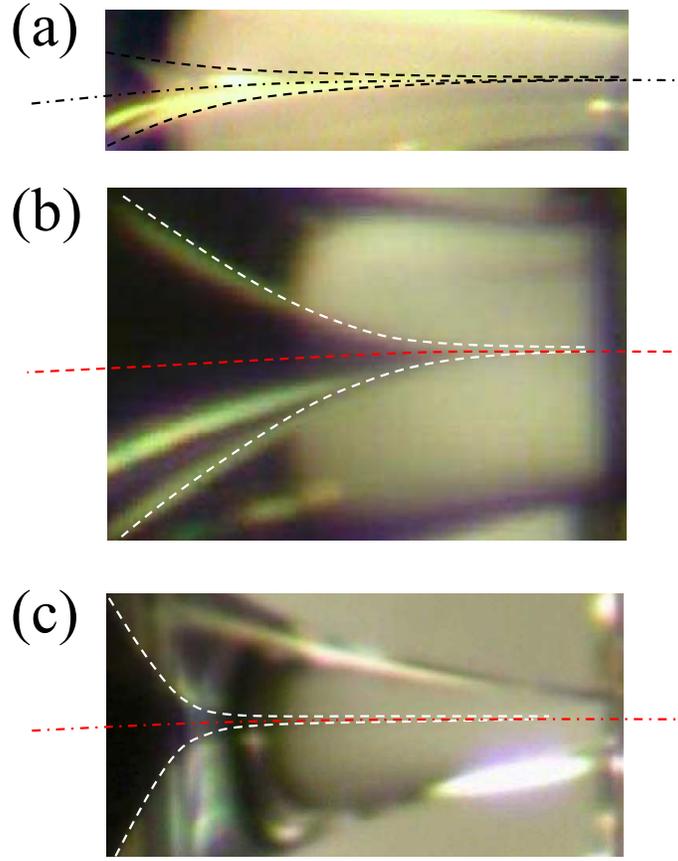}}
\caption{(Color online) Different coflowing capillary jets into a much thicker silicone oil (which
is in turn focused by an outer gas flow), from Ref \cite{Gan07NatPhys}: (a) an air jet
($\mu=2.06\times 10^4$); (b) a mercury jet ($\mu=2.42\times 10^2$); (c) an ink solution jet
($\mu=22.2$). The lines guide the eye along the jet interface. Dash-dotted lines approximately
delineate the positions of the centers of the cross section area, not in a straight line owing to a
slight bend of the outer flow. Observe that the overall jet slope increases as the viscosity ratio
$\mu$ decreases.} \label{jet}
\end{figure}

\begin{figure}
\centerline{\includegraphics[width=0.85\textwidth]{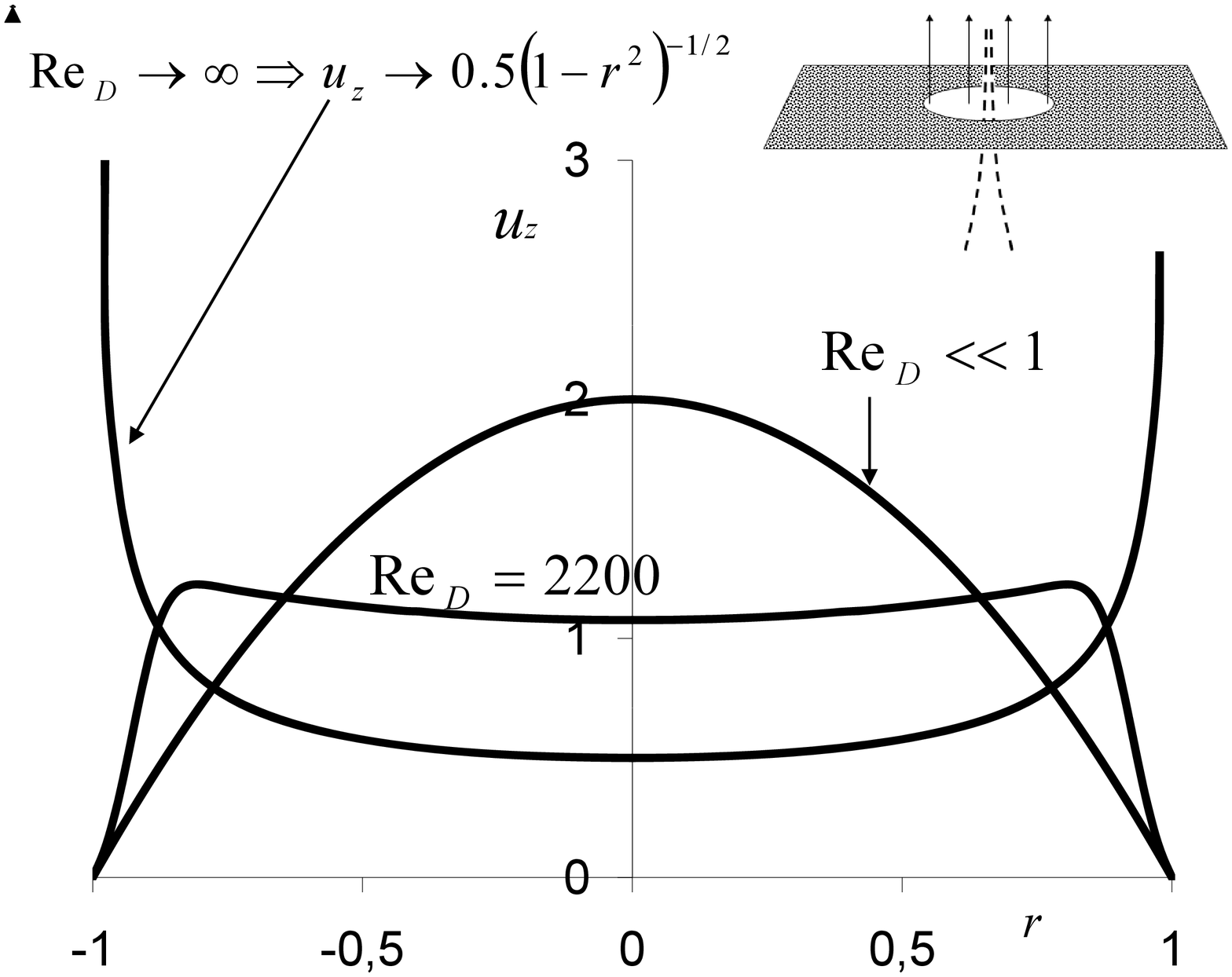}} \caption{Axial velocity profile
$u_z=U_z/\bar{U}_2$ as a function of the radial coordinate $r$ (made non-dimensional with $D/2$)
for an incompressible fluid at a round orifice in an infinite plane thin wall as a function of the
radial coordinate for various ${\text Re}_D$ values. The flow is forced through the orifice by a
pressure difference across the thin wall. In particular, the case ${\text Re}_D=2200$ has been
obtained using numerical simulation (Volumes of Fluid). The inset shows the basic flow
configuration.}\label{OF}
\end{figure}

The link connecting convective/absolute instability to jetting/dripping, respectively, is well
documented by experiments \citep{Lin2003,Sev05,AP06,Guillot2007}. Thus, the rate at which the
perturbations grow is a function of ${\text Re}$ and $Ca$. Since the jet disperses fragments of
fluid 1 into an immiscible jet 2 of density $\rho_2$ and viscosity $\mu_2$, two additional
fundamental parameters are the fluid density and viscosity ratios $\rho=\rho_2/\rho_1$ and
$\mu=\mu_2/\mu_1$. When jetting is produced by flow focusing\cite{Gan98}, one has a stream of fluid
2 forced through an orifice of diameter $D$ which focuses the jet of fluid 1 (see Figure 2, inset).
In flow focusing, we assume that $R\ll D$, the most usual case. The relationship between the axial
velocity $U_z$ of fluid 2 near the axis of the exit orifice, $U_z=U_2$, and the flow rate of fluid
2 through the orifice, $Q_2$, depends on the Reynolds number ${\text Re}_D=\rho_2 Q_2/(\mu_2 D)$
(see Figure \ref{OF}). For ${\text Re}_D\rightarrow \infty$, one asymptotically has\cite{MF53}
$U_2=0.5 \bar{U}_2$, where $\bar{U}_2=4Q_2/(\pi D^2)$. However, for ${\text Re}_D\ll 1$, one
approximately has $U_2\simeq 2\bar{U}_2$. For moderate to high ${\text Re}_D$, a useful
approximation is $U_2\sim \bar{U}_2$ (see Figure \ref{OF} for ${\text Re}_D=2200$).
Besides, when viscous effects dominate in the flow of fluid 1, the jet dynamics is thus determined
by the parameters $\{Ca,\rho,\mu\}$ only, since ${\text Re}$ disappears from the analysis. In the
limit ${\text Re}\rightarrow\infty$, ${\text Re}$ is out from the analysis as well, and viscous
effects are confined to boundary layers at the interface\cite{Gan07PRE}. Since in this case the jet
radius $R$ is implicitly given by the equation $\rho_2 U_2^2=2\sigma/R+\rho_1 U_1^2$, the jet
dynamics becomes governed by $\{W\!e,\rho,\mu,U\}$, where $U=U_2/U_1$ and $U_1=Q_1/(\pi R^2)$
($Q_1$ is the issued liquid flow rate of fluid 1). These two limiting cases allow for analytical
treatment, and will be discussed in this work.

\begin{figure}
\centerline{\includegraphics[width=0.75\textwidth]{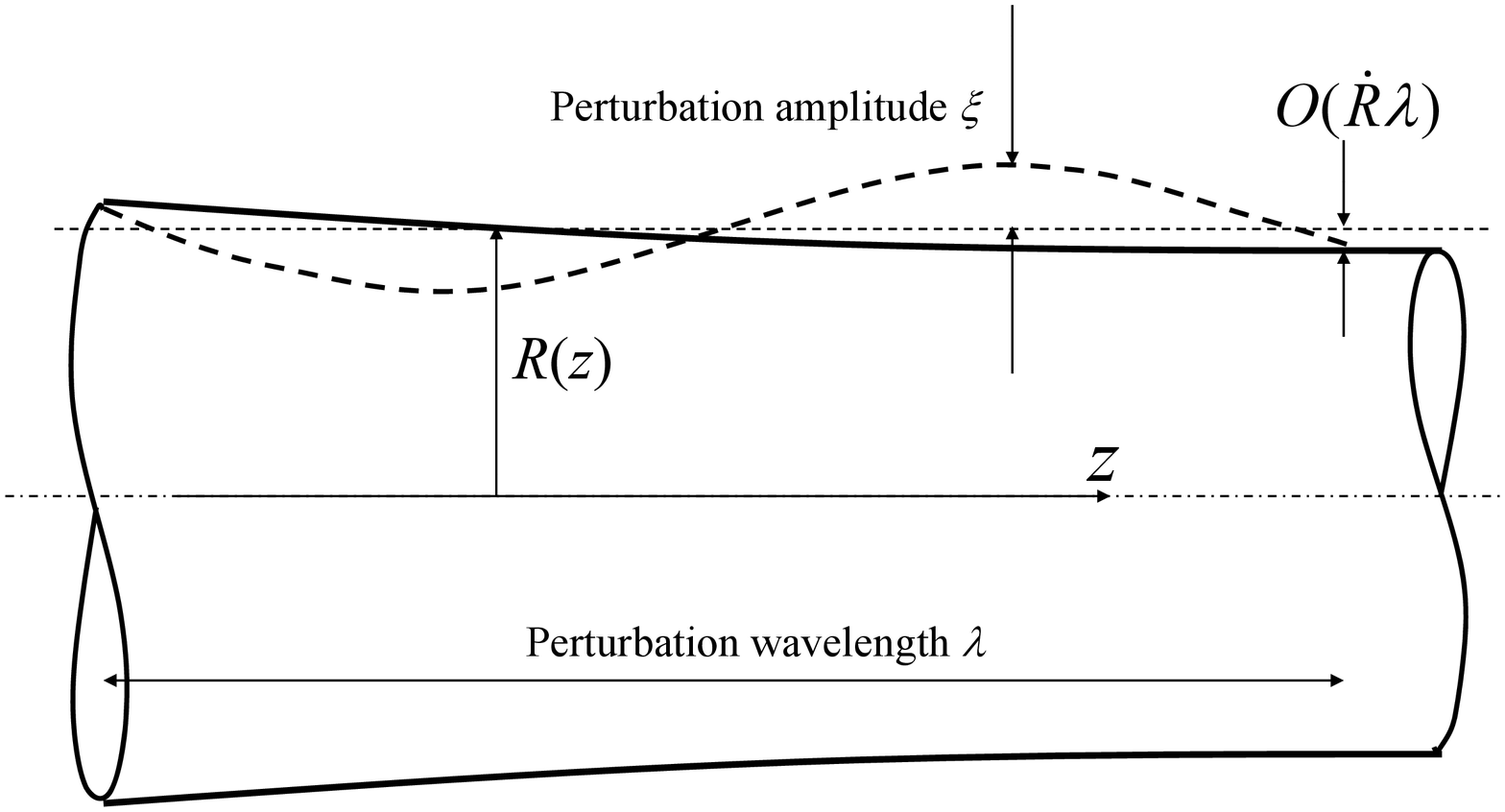}} \caption{Sketch of a perturbed jet
portion.}\label{fig-lsh}
\end{figure}

Naturally, a jet is characterized by its slenderness and the applicability of the usual slender
flow approximations. To study the linear stability of the system, we interrogate the jet response
to small perturbation amplitudes $\xi\ll R$ with a given axial wavelength $\lambda$ (see figure
\ref{fig-lsh}).
As long as the jet curvature is very approximately $R^{-1}(z)$, and the jet radius variations along
axial distances $\lambda$, of the order of $O(\dot{R}\lambda)$ (where $\dot{R}=\fr{d R}{d z}$), are
small compared to $R$, one can always choose amplitudes such that $O(\dot{R}\lambda)\ll O(\xi)\ll
O(R)$, which justifies the classical cylindrical approximation. Under these assumptions, we
investigate the linear dynamics of cylindrical steady capillary jets in two limits: (i) creeping
flow limit (${\text Re}\rightarrow 0$), and (ii) inertia dominated limit (large ${\text Re}$). In
the first limit, our exploration will focus on some practical scenarios of jetting at very small
scales very recently reported in the
literature\cite{Anna03,SB2006,CdPE2006,AnnaMayer06,Guillot2007,Gan07NatPhys}.

Consider a thin cylindrical jet of fluid 1, moving in a co-flowing fluid 2. The general capillary
flow category claimed in this work, for which the flow rate $Q_1$ could be made arbitrarily small
(i.e., jetting would always take place independently of the issued flow rate), requires either
making the jet radius $R$ very small or allowing for recirculation (part of the jet inner flow
should flow upstream) for continuity. Asymptotically, one may as well have a static jet core
($U_1=0$) surrounded by a co-flowing fluid with velocity $U_2$, which demands a negligible viscous
diffusion from the interface. In this work, we will study these configurations which allow analytic
dispersion relations. Thus, the occurrence of UJ is not restricted to either small or large ${\text
Re}$ values: in fact, the only necessary condition is that the jet surface velocity should be large
enough to convect all perturbations downstream. The limit cases of large and small ${\text Re}$
will be discussed in some detail in this work because they allow for analytical treatment and for
their importance and illustrative power. In this sense, given the utter importance of controlled
micron- and nano-sized droplet generation, we aim to provide a global understanding of mechanisms
supporting UJ, to guide future fluid disperser designs of special relevance in chemical
engineering, combustion and energy efficiency, transport, food processing, spraying, biochemistry,
pharmacy, biomedicine, environmental engineering, among others.

A steady capillary jet of a fluid surrounded by an immiscible continuum fluid phase (figure
\ref{jet}), an intrinsically unstable state\cite{Powers-Goldstein97}, is locally stable (jetting is
possible) whenever the so called marginal stability velocity $v^*_-$ relative to an observer is
positive, so that all perturbations are convected downstream. From now on, whenever we assume a \sl
local \rm analysis applicable, velocities made non-dimensional with the local surface velocity
$U_s$. Since dripping is axisymmetric, here we consider axisymmetric perturbations only. Assuming
small perturbations as superposition of waves proportional to $\exp[i(kz-\omega t)]$, where wave
number $k=k_r+ik_i$ and frequency $\omega=\omega_r+i\omega_i$ (made non-dimensional with $R^{-1}$
and $U_s/R$, respectively) are complex numbers, the jet linear dynamics is governed by the
dispersion relation between $k$ and $\omega$. Central to our analysis is the fact that steady
capillary jets are unstable states\cite{Saarloos88,Powers-Goldstein97}, that is, they exhibit wave
number ranges $k$ such that perturbations grow in time as $\exp[\omega_i t]$ leading to break up. A
growing disturbance usually spreads along the jet bounded by two fronts moving with velocities
$v^*_+$ and $v^*_-$, the extremal values of the envelope velocities $v=\omega_i/k_i$
\citep{Dee-Langer83,Saarloos87,Saarloos88,Powers-Goldstein97}. The extremal values or marginal
stability velocities $v^*$ should satisfy \be v^*=\omega_i^*/k_i^*=\partial \omega_i/\partial
k_i|_{k=k^*}\,,\quad \partial \omega_i/\partial k_r|_{k=k^*}=0.\label{msv}\ee

Thus, the fate of the jet at a source-fixed station is determined by the minimum marginal stability
velocity $v^*_-$. If $v^*_->0$, small perturbations are convected downstream for all wave numbers
(convective instability, or local stability), while if $v^*_-<0$, some wave number ranges will grow
locally without bound (absolute instability). The aim of this work is to report a special class of
parametric conditions of capillary jetting for which the marginal stability velocity $v^*_-$
(minimum front propagation velocity) keeps {\it always positive} for vanishing dispersed flow rates
of fluid 1. We designate this rather singular flow condition ``unconditional jetting'' (UJ). This
means that the capillary jet is {\it convectively unstable}, or {\it locally stable}, and does not
undergo a jetting-dripping transition as the issued flow rate $Q_1$ vanishes. The technological
relevance of this class of flows can be understood as follows: picture a steady capillary jet
flowing down from a slightly opened tap. If $v^*_-$ were always positive, one could slowly turn off
the tap without transition to dripping. Eventually, when the tap is turned off, the jet would thin
down to the continuum limit without transition to dripping. For extremely small flow rates, nearly
monodisperse droplets of any imaginably small size would be produced upon Rayleigh breakup at a
highly controllable rate. Although this behavior is indeed rather unusual for laminar jets from
taps, it is however a real occurrence in co-flowing jets for a certain rather ample ratios of
continuous-dispersed fluid densities and viscosities\cite{Gan07} $\rho$ and $\mu$, or when the
continuous phase 2 co-flows with the jet at a velocity larger than a critical
velocity\cite{Gan07PRE,Gan07NatPhys} $u^*_2$. We will see that the fundamental physical requirement
for UJ is to have a jet surface velocity above a critical one.

\section{The case of negligible inertia: ultra-thin jetting}

First, we will consider the case where velocities of both fluids are equal to the surface velocity
(flat velocity profiles). Here, the jet's linear dynamics is governed by the following dispersion
relation \cite{AP06,Gan06}: \be i Ca(\omega-k)\left[\fr{N(k,\omega,{\text
Re},\rho,\mu)}{D(k,\omega,{\text Re},\rho,\mu)}+ 2(1-\mu)\right]+ (k^2-1)= 0 .\label{dr1}\ee For
convenience, we define ``viscous'' wave numbers for both fluids as: \be k_1^2=k^2-i\,{\text
Re}\cdot(\omega-k)\,,\quad k_2^2=k^2-i\rho\,\mu^{-1}\,{\text Re}\cdot(\omega-k),\ee Using these
definitions, functions $N$ and $D$ are expressed as:
\begin{eqnarray}
N\equiv 2 k \mu k_1 k_2 \left[K_0(k_2) I_1(k_1) k_1 + I_0(k_1)
K_1(k_2) k_2\right] \nonumber\\
+k\left[k^2(\mu-1)-k_1^2+\mu k_2^2\right]^2 I_0(k) I_1(k_1) K_0(k) K_1(k_2) \nonumber\\
+4 k^3 k_1 k_2 (\mu-1)^2 I_0(k_1)I_1(k)K_0(k_2)K_1(k) \nonumber\\ -k_2
I_1(k_1)K_0(k_2)\left\{\left[k^4+k_1^2k_2^2+k^2(k_1^2-k_2^2)\right]\mu
I_1(k)K_0(k)  \right. \nonumber\\
\left.  +\left[k_1^4+k^4(1-2\mu)^2-2k^2 k_1^2(\mu-1)\right]I_0(k)K_1(k)\right\}\nonumber\\ -k_1
I_0(k_1)K_1(k_2) \left\{\left[k^4(\mu-2)^2+2k^2k_2^2\mu(\mu-1)+\mu^2k_2^4\right]
I_1(k)K_0(k) \right. \nonumber\\
+\left. \left[k^2(k^2-k_1^2)+k_2^2(k^2+k_1^2)\right]\mu I_0(k) K_1(k)\right\}
\end{eqnarray}
\begin{eqnarray}
D\equiv k\left\{\left[k_2 K_0(k_2) K_1(k)- k K_0(k) K_1(k_2)\right](k_1^2-k^2) I_1(k) I_1(k_1) +
\right.
\nonumber\\
\left. \mu\left[k_1 I_0(k_1) I_1(k)-k I_0(k) I_1(k_1)\right](k_2^2-k^2)K_1(k) K_1(k_2)\right\}
\end{eqnarray}

Interestingly, a series expansion of $N$ and $D$ around ${\text Re}=0$ yields \be
N=\fr{(\omega-k)^2}{4 k}N_2(k,\mu){\text Re}^2+O({\text Re}^3)\,,\quad D=\fr{(\omega-k)^2}{4\mu
k}D_2(k,\mu){\text Re}^2+O({\text Re}^3),\ee where $N_2$ and $D_2$ (omitted, lengthy expressions)
are independent of $\omega$ and $\rho$, as can be checked using Mathematica\circledR. Moreover, in
this limit the fluid velocities \sl do not need to be uniform\rm. Defining for convenience an
average capillary number as \be\bar{Ca}=\mu^{1/2}Ca=(\mu_1\mu_2)^{1/2} U_s/\sigma,\label{avca}\ee
equation \ref{dr1} yields an explicit analytical expression for $\omega=\omega(k,\bar{Ca},\mu)$
providing a closed form for the frequency $\omega$: \be
\omega=k+i(k^2-1)\bar{Ca}^{-1}\left[\mu^{1/2}\fr{N_2(k,\mu)}{D_2(k,\mu)}+
2(\mu^{-1/2}-\mu^{1/2})\right]^{-1}.\label{v}\ee This equation, together with conditions \ref{msv},
provides $v^*$, $\omega^*$ and $k^*$ for a given set $\{\bar{Ca},\mu\}$. Since $\bar{Ca}$ does not
depend on the jet diameter $d$, a particularly strong practical implication follows: if one can
find a parametrical region $\{\bar{Ca},\mu\}$ where $v^*$ is always positive, it is so for any
value of the jet diameter, no matter how small it can be.

\begin{figure}[htb]\centerline{\includegraphics[width=0.75\textwidth]{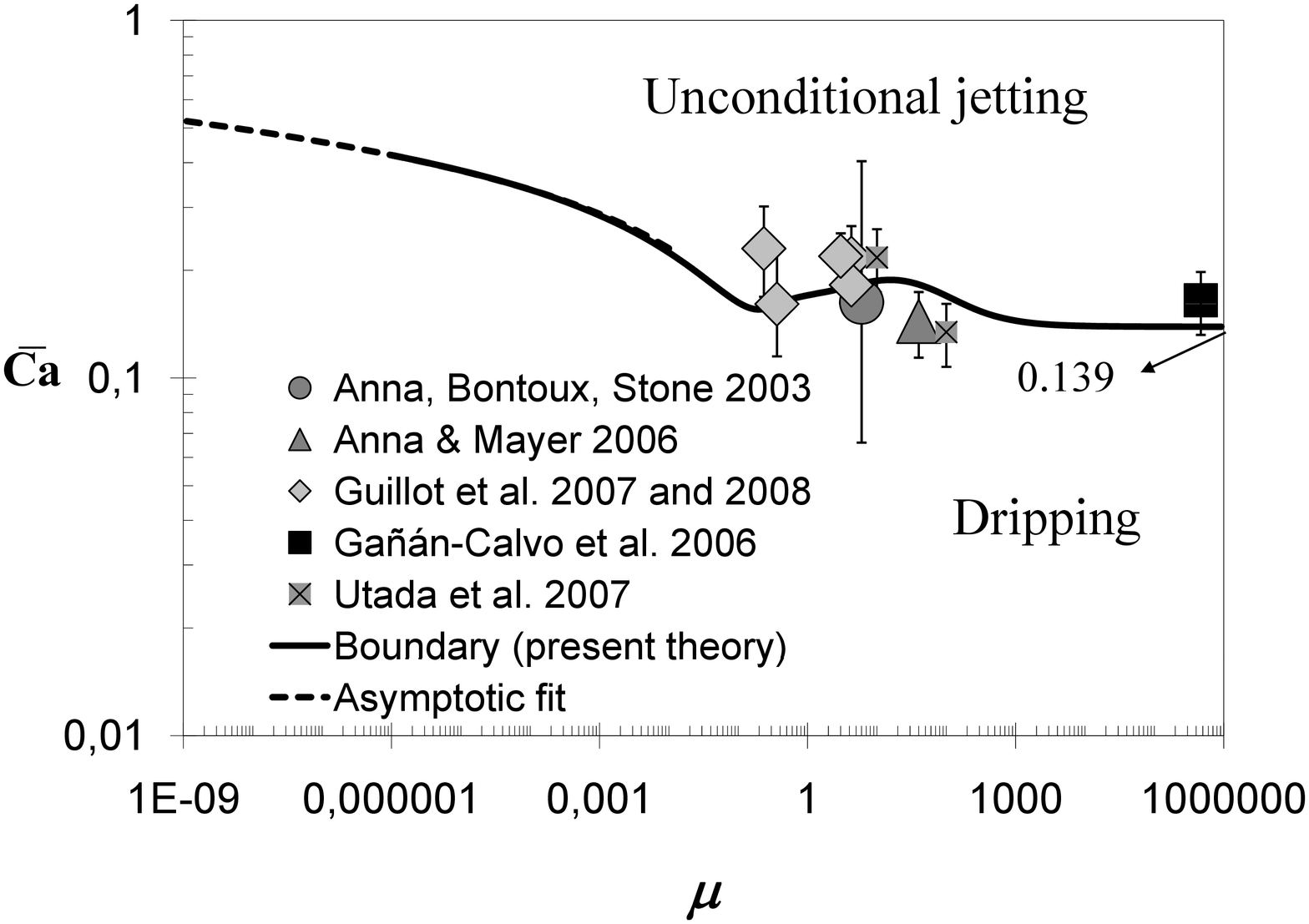}}
\caption{Critical Capillary number as a function of the viscosity ratio $\mu$. An asymptotic fit
for $\mu\ll 1$ is provided: $\bar{Ca}^*\rightarrow 0.1[\ln(\mu)]^{0.5465}$. Also given: comparison
with published experimental results were conditions for observed dripping/jetting transition are
recorded, assuming a homogeneous velocity profile of both fluids in and out the jet. Unless
otherwise stated, the estimated errors associated to data extraction from published plots are about
$\pm 20$\%. Data taken from (symbols, authors, data source figures): $\bigcirc$ Anna \it et al. \rm
\cite{Anna03}, Fig 3; $\bigtriangleup$ Anna and Mayer\cite{AnnaMayer06}, Fig. 5a; $\times$ Utada
\it et al.\rm\cite{Utada2007}, Fig. 4; $\diamond$ Guillot \it et al. \rm\cite{Guillot2007}, Figs.
5(a1,a2), and Herrada, Ga\~n\'an-Calvo \& Guillot 2008 \cite{HGCG2008}. Fig. 3; $\square$
Ga{\~n\'a}n-Calvo \it et al. \rm\cite{Gan06}, Fig. 4. (liquid surface tension was approximately 50
mN/m; liquid velocity at the orifice entrance can be calculated approximately\cite{Gan2004EPJB} as
$U= k \Delta P\cdot D/\mu_2$, where $\Delta P$ is the pressure drop through the orifice, and $k$ is
about 0.5). In particular, for Anna \it et al. \rm \cite{Anna03}, $\mu_1=0.001$ Pa$\cdot$s,
$\mu_2=0.006$ Pa$\cdot$s, 0.005$<\sigma<$0.01 N/m (personal communication from the authors),
1.4$<Q_2<$4.2 $\mu$L/s or 0.27$<U_2<$0.83 m/s, which gives 0.066$<\bar{Ca}<$0.404 in this case.}
\label{f2}
\end{figure}
Figure \ref{f2} shows a plot of the loci $\bar{Ca}=\bar{Ca}^*(\mu)$ where $v^*=0$. Fifteen orders
of magnitude in $\mu$ are explored, showing a small dependency of the critical $\bar{Ca}$ on $\mu$,
which supports our definition choice for a relevant capillary number in the creeping flow limit
(jet radii $R\ll \mu_1/(\rho_1 U_s)$), which incorporates \it both \rm inner and outer fluid
viscosities. This curve splits the $\{\mu,\bar{Ca}\}$ plane in two halves: above(below) this curve,
$v^*_-$ is always positive(negative) and the jet flow is always supercritical(subcritical)
independently of the jet diameter (supercritical: convective velocity always overcomes the upstream
spreading of perturbations). Thus, if the velocity profile of both fluids is homogeneous in and out
of the jet, supercritical jets of any imaginably small diameter could be produced for a co-flow
speed $u_s$ larger than a critical velocity $u_s^*=\bar{Ca}^*$ ($U_s^*=\sigma
\bar{Ca}^*/(\mu_1\mu_2)^{1/2}$).

Experimental results of other authors are compared with theory in Figure \ref{f2}. Anna et
al.\cite{Anna03} used a planar flow focusing device where they dispersed water ($\mu_1=1$
mPa$\cdot$s, $\rho_1=1$ kg L$^{-1}$) in silicone oil ($\mu_2=5$ mPa$\cdot$s, $\rho_2\simeq 0.9$ kg
L$^{-1}$). Their experiments show (Anna et al.\cite{Anna03}, Figures 3e,k,q) that jetting was found
for values of the focusing oil flow rate $Q_2=4.2\,\mu L$s$^{-1}$ and above. Since their Reynolds
number at the orifice was about 6, a calculation of the oil velocity at the orifice axis yields
about $U_2=2 Q_o/(h D)=1.65$ m/s, where $h=117\,\mu$m and $D=43.5\,\mu$m are the orifice depth and
width, respectively (their orifice length was about $L=120\,\mu$m from their pictures, and thus a
parabolic velocity profile should have developed). In accord with our predictions, they found
jetting to occur above the indicated oil flow rate independently of the oil-water flow rate ratio,
i.e. \it independently of how thin the jet was\rm. Their corresponding threshold
$\bar{Ca}=(\mu_1\mu_2)^{1/2}U_2/\sigma$ is about 0.169 (with errors associated to surface
tension\cite{Anna2007personal} and indetermination between $Q_o=1.4\, \mu$L/s and 4.2 $\mu$L/s).
Besides, using their same planar flow focusing device, Anna and Mayer\cite{AnnaMayer06} recently
reported jetting \it independently \rm of the focused flow rate beyond a capillary number
$\bar{Ca}=0.144$ (worked out from their disclosed data), for an aqueous solution focused by oil
with $\mu=40$, in the absence of surfactants. They reported transition from dripping to jetting for
outer-to-inner flow rates ratios as large as 300, corresponding to jet diameters as small as about
3 $\mu$m. When surfactants are present, their results cannot be compared owing to non-linear
dynamic surface tension effects beyond the critical micelle concentration c.m.c. (as they declare)
at the jet.

Moreover, Guillot et al.\cite{Guillot2007} have reported an extensive and very valuable series of
experiments of a liquid jet flowing coaxially in another immiscible liquid inside a cylindrical
channel. When the jet to channel diameter ratio becomes very small, their measurements can be
compared to our predictions. Their results show the dripping to jetting transition for a viscosity
ratio$\mu=4.3$ at $\bar{Ca}=0.18$ and 0.22 (calculated from their published data in their figures
a1 and a2, respectively, for their higher available outer-to-inner flow rates ratios). We do not
make use of their results with surfactants for the same reasons above given upon results from Anna
and Mayer\cite{AnnaMayer06}. Finally, for very large outer-to-inner viscosity fluid ratio,
$\mu=4.7\times 10^5$ (syrup-air), we have found\cite{Gan06} transition from bubbling to jetting at
$\bar{Ca}\simeq 0.18$. All these experimental findings, plotted in Figure \ref{f2}, provide full
support to our prediction for homogeneous (flat) velocity profiles.

\begin{figure}
\centerline{\includegraphics[width=0.85\textwidth]{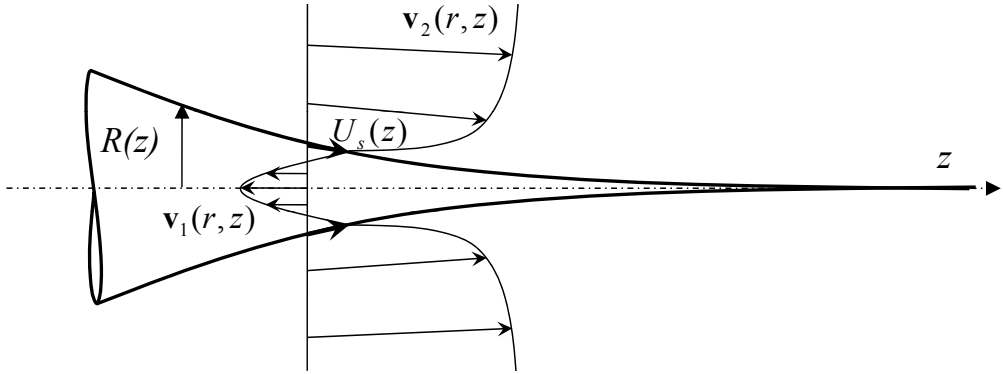}} \caption{Sketch of a jet tapering
from a finite sized source in the form of an arbitrarily thin spout. In this configuration, the
fluid velocities cannot be homogeneous for conservation of mass.}\label{fig-jet2}
\end{figure}

\begin{figure}
\centerline{\includegraphics[width=0.75\textwidth]{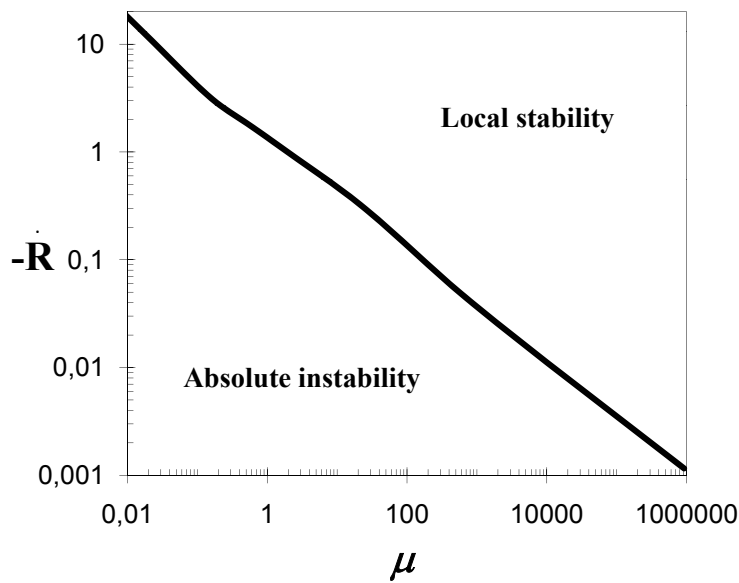}} \caption{Plot of the limiting slope
of the jet as a function of the viscosity ratio $\mu$.}\label{dr}
\end{figure}

Furthermore, we will consider a limit situation where the jet tappers from a finitely sized source
(e.g. a capillary tube of inner radius $R_o$), forming an arbitrarily thin spout of fluid 1 (figure
\ref{fig-jet2}). For ease of understanding in this case, where we need to consider axial dependency
of jet radius and velocities, we turn to dimensional expressions. Here, the local basic velocity
profiles of fluids for negligible inertia, satisfying stress balance at the jet's surface, are
given by:
\begin{eqnarray}
U_1(r,z)=U_s(z)+\fr{\sigma \dot{R}(z)}{4 \mu_1}\left[1-\left(\fr{r}{R(z)}\right)^2\right]\nonumber \\
U_2(r,z)=U_s(z)-\fr{\sigma \dot{R}(z)}{2
\mu_2}\log\left(\fr{r}{R(z)}\right)\label{profiles}\end{eqnarray} To obtain these approximate
equations, we have assumed that (i) the jet is slender ($\dot{R}$ is sufficiently small) and
transversal velocities are neglected, and (ii) the outer pressure becomes negligible compared to
the inner jet pressure $p_1\simeq \sigma/R(z)$ as the spout radius becomes very small.
Now, mass continuity for a vanishing issued flow rate of fluid 1, $Q_1\simeq 0$, yields: \be
U_s(z)=\fr{-\sigma \dot{R}(z)}{8\mu_1}\ee Substituting this value of the surface velocity in the
expression \ref{avca} of the average capillary number, one has that the jet would be \it locally
stable \rm if its \it local \rm slope satisfies: \be -\dot{R}<\fr{8
\bar{Ca}^*}{\mu^{1/2}}\label{rd}\ee This limiting slope is only a function of the viscosity ratio
$\mu$, plotted in figure \ref{dr}.
To obtain function $\dot{R}^*=\dot{R}^*(\mu)$, we use the fundamental fact that the critical
capillary number is independent of the velocity profile shape when ${\text Re}\ll 1$, plugging in
the values of the critical capillary number of figure \ref{f2}. This reveals that the jet slope is
limited by the inverse of the viscosity ratio $\mu$ only, once the flow of both fluids is dominated
by viscosity. It is worth noting that viscosity ratios smaller than about 0.01 would always
provide, for ${\text Re}\ll 1$, a \it local \rm stability of the tapering meniscus if the flow
configuration allows co-flowing speeds $U_s$ larger than $\sigma \bar{Ca}^*/(\mu_1\mu_2)^{1/2}$,
once jet is issued.

The latter result, asymptotically valid for $\dot{R}$ small, is of fundamental importance to at
least qualitatively explain experimental observations. In fact, if one observes figure \ref{jet},
the overall spout slope decreases as the viscosity ratio $\mu$ increases, according to equation
(\ref{rd}). Moreover, it also explains why gas spouts are extremely difficult to achieve in
coflowing liquids, since the local requirement (\ref{rd}) becomes very difficult to fulfill unless
a very small gas source is used. Nevertheless, upstream of the tapering thin spout, when the
meniscus slope becomes of the order unity, the local Reynolds number may not be necessarily small
and the limiting conditions for local stability deviate significantly from the above requirements.
In fact, the critical capillary number \it decreases \rm when Re increases (see Montanero \&
Ga\~{n}\'{a}n-Calvo\cite{Montanero08}, figure 5), and therefore the critical slope may increase, becoming
of the order unity, in accord with real configurations like the ones shown in figure \ref{jet}.
Notwithstanding this, again, requirement (\ref{rd}) gives for the first time at least a qualitative
explanation to observations on the slopes of tapering spouts and the difficulty to obtain stable
gas spouts.

A remark on the experiments by Courrech du Pont and Eggers\cite{CdPE2006} is here necessary: these
authors report a ``universal'' critical value of the slope $\dot{R}=0.47\pm 0.06$ of a tip
singularity in a viscous combined withdrawal of air through a small orifice by means of a viscous
liquid, a silicone oil with $\mu_2=30$ or $60 Pa\cdot s$. This remarkable result, from a large set
of experiments, points to the existence of a locally self-similar conical region which eventually
should tapper into an arbitrarily thin gas spout, where condition \ref{rd} should hold once the gas
flow becomes viscosity-dominated in the spout. That self-similar region (we do not deal with this
problem here) should be a function of the viscosity ratio $\mu$ only, although the structure of
that locally self-similar flow is not known yet. However, our theory predicts critical slopes
(about $6\times 10^{-4}$) much smaller than the one reported in \cite{CdPE2006}, should a sustained
(steady) thin spout be formed. Here we propose that the flow observed in \cite{CdPE2006} does not
actually yield a \it steady \rm extremely thin jet downstream of the observed region, within the
suction orifice, for the experimental condition tested. In fact, these authors report that their
setup ``fails to produce a thin spout'', and that visible bubbles appear downstream of the orifice
when entrainment takes place. The entrained spout would only be locally stable down to any
imaginably small scale, leading to bubble sizes comparable to the spout diameter, if they had a
local velocity at the orifice larger than 0.312 m/s or 0.441 m/s for their experimental conditions
($\mu_2=60$ or 30 Pa$\cdot$ s, respectively, with $\sigma=0.0213$ N/m and $\bar{Ca}^*=0.139$).
Since their velocities were below 0.0254 m/s (velocity at the axis corresponding to a flow rate of
0.01 ml/s through an orifice of 1 mm for viscosity dominated flow) they could not form a sustained
extremely thin spout according to our prediction. However, the molecular mean free path for air
enters into play below 1 micrometer \cite{CdPE2006}, and our theory becomes questionable below that
scale.

Finally, the elegant mathematical solution for a steady thin spout of Zhang \cite{WWZ2004} would be
stable if her local convective velocities at the spout were everywhere larger than the critical
ones here predicted. This condition would add an extra requirement in her analysis.

The implication of present conclusions in microfluidics and, in particular, in the field of device
design for emulsification are highly attractive for technological application.

\section{A case with dominant inertia.}

\begin{figure}[htb]\centerline{\includegraphics[width=0.55\textwidth]{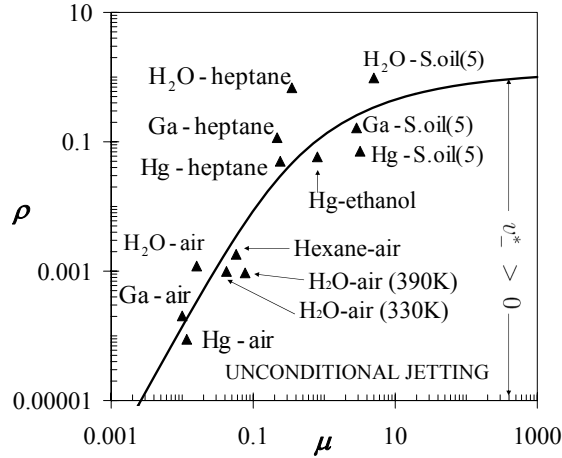}}
\caption{Parametrical region in the $\{\mu,\rho\}$ space where unconditional jetting is
found for flow focusing. Thirteen fluid combinations of interest are shown.
Interestingly, water focused by air can theoretically exhibit unconditional jetting for
ambient temperatures above $320K$.}\label{f3}
\end{figure}

While the above given analysis is restricted to negligible inertia, in the limit of dominant
inertia one can also find an interesting, analytically tractable asymptotic case of UJ
\cite{Gan07}, i. e. where the issued flow rate $Q_1$ can be made arbitrarily small as well. In this
alternative case, viscous effects are confined to thin boundary layers at both sides of the jet
surface, where bulk velocities $U_1$ and $U_2$ are nearly flat but different. If both layers
develop simultaneously from the same station near the dispersed fluid source, while they are thin
compared to the jet diameter the surface velocity $U_s$ can be explicitly expressed as (see the
arguments given in \cite{Gan07}): \be
u_s=U_s/U_1=\fr{1+(\rho\mu)^{1/3}U}{1+(\rho\mu)^{1/3}},\label{us}\ee where, again, $U=U_2/U_1$. In
this situation, if boundary layers are sufficiently small compared to the jet radius, the
dispersion relation for an infinite cylindrical capillary jet moving with bulk velocity $U_1$ in a
co-flowing fluid with bulk velocity $U_2$ is \be (\omega -k)(\omega-k
u_s)\fr{I_0(k)}{I_1(k)}+\rho(\omega-k u_s)(\omega-k U)\fr{K_0(k)}{K_1(k)}=\fr{\rho U^2
k(k^2-1)}{W\!e_2},\label{dr2}\ee where $W\!e_2=\rho_2U^2_2 d/(2\sigma)$. Note that the Reynolds
number is absent in this expression, consistently with the assumption of dominant inertia. Viscous
effects are however important at the jet surface, making $\mu$ fundamentally important through the
jet surface velocity $u_s=u_s(U,\rho,\mu)\neq 1$. If we seek conditions yielding small dispersed
flow rates, i.e. $U\rightarrow \infty$, equation (\ref{dr2}) plus conditions \ref{avca} with
$v^*_-=0$ yields the critical value $W\!e_2^*$ using expansions $\omega=\omega_0+U^{-1}\omega_1$
and $k=k_0+U^{-1}k_1$ (see note\cite{note3}). Some flow configurations, such as flow focusing, give
rise to further constraints; for very low viscosities, equilibrium requires $W\!e_2\rightarrow 2$
for $U\rightarrow \infty$ (from equation $\rho_2 U_2^2=2\sigma/R+\rho_1 U_1^2$). In this case, for
a given density ratio $\rho$, supercritical conditions are found for continuous phase viscosity
larger than a critical ratio over the dispersed phase viscosity, or when $\rho$ is smaller than a
critical ratio for a given $\mu$ value, as shown in figure \ref{f3} (see also \cite{Gan07}, Figure
3) where thirteen possible flow focusing combinations are plotted. Note that the points correspond
to given fluid properties, not reflecting transitions: the fluid combinations in the region
``unconditional jetting'' would exhibit this behavior, while those outside that region would only
show dripping/jetting transition for a non-zero flow rate $Q_1$. In other words, for the points in
the region of UJ in figure \ref{f3}, surface velocities are always large enough to have $v^*_->0$
and perturbations are flushed downstream, even though the bulk fluid can be literally static.

\section{Concluding remarks}

\begin{figure}[htb]\centerline{\includegraphics[width=0.55\textwidth]{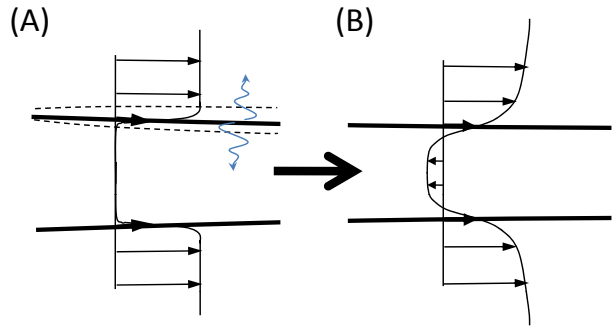}}
\caption{Illustrative sketch of the UJ case for dominant inertia. (A) Flow within length $L_\mu$
from the jet source, where the asymptotic analysis is valid but viscous momentum diffusion from the
surface thickens the boundary layers; (B) Development of a recirculation pattern downstream of the
length $L_\mu$ owing to viscous diffusion.}\label{final}
\end{figure}

Interestingly enough, upon consideration of Ga{\~n\'a}n-Calvo\cite{Gan07} equation (8), the limit
$U\rightarrow \infty$ here considered is strictly valid independently of the jet electrification as
well, since the electrical term (finite) is overcome by other terms proportional to $U\gg 1$ and
$U^2$. Obviously, viscous diffusion of momentum from their surface soon affect the entire cross
section of these jets. From well known boundary layer analysis, the downstream axial length $L_\mu$
that must be traveled to achieve $\delta_1\sim d$ should satisfy $(\mu_1 L_\mu
\rho_1^{-1}U_s^{-1})^{1/2}\sim\delta_1\sim d$. Thus, using equation (\ref{us}), the maximum
``unconditionally stable'' length $L_\mu$ is limited to \be L_\mu/d \sim \fr{\sigma}{\mu_1
U_2}\fr{(\mu\rho)^{1/3}}{\rho[1+(\mu\rho)^{1/3}]}.\ee The equation for the surface velocity
(\ref{us}) is then only valid within a length $L_\mu$ from the jet source, but not downstream.
After this length, one cannot use the dispersion relation \ref{dr2} since the bulk velocities
become inhomogeneous; in fact, the liquid bulk should become recirculating in order to accomplish
the vanishing $Q_1$ condition (see figure \ref{final}).
Naturally, other cases of UJ may exist when inertia and viscous forces are comparable [e.g. that
sketched in Fig. 8 (B)], but their required numerical treatment puts them outside the scope of the
present analysis.

As a final remark, it is worth stressing the fundamental importance of the surface velocity $U_s$
in the present analysis, for both creeping and inertia-dominated flows: note that for the cases
here studied, where analytical solutions can be found, the wave behavior with respect to an
observer is strictly determined by the linearized kinematic condition at the interface: \be
\fr{\partial R}{\partial t} + U_s\fr{\partial R}{\partial z} -V=0,\ee where $V$ is the small
transversal component of the surface velocity. This is the only equation where the surface velocity
appears: interestingly, since the first order velocity profile does not enter in the analytical
dispersion relations here studied \cite{note2}, when the observer moves with the surface, its
velocity disappears from the analysis, and one recovers the classical dispersion relations for
cylindrical capillary columns existing in the literature.

This work is supported by the Ministry of Science and Technology of Spain, grant no. DPI2004-07197.
Suggestions from Dr. Pascual Riesco-Chueca and discussions with Dr. Jos\'e M. Montanero are highly
appreciated. Dr. Miguel A. Herrada kindly supplied the unpublished data for the case ${\text
Re}_D=2200$ in figure \ref{OF}.


\begin{thebibliography}{28}
\bibitem[not()]{note3}
Complex equation (\ref{dr2}) provides six scalar equations from its real and
  imaginary terms of order O(1), O($U$) and O($U^2$). Conditions for $v^*_-=0$
  provide three scalar equations resulting in Im$(\omega_0)=0$,
  Real$(\omega_1)=0$ and Im$(\omega_1)=0$.
\bibitem[not()]{note2}
In the viscosity dominated flow, the perturbed equations for both the pressure and the velocity are
independent of the steady, first order velocities (the convective term, the one responsible for a
possible coupling between first order and perturbation, is absent). In the inertia-dominated flow
here considered, since the basic velocity profiles are homogeneous, they only enter the analysis
through the constant parameter $U=U_2/U_1$.
\end{thebibliography}
\end{document}